# Generating arbitrary polarization states by manipulating the thicknesses of a pair of uniaxial birefringent plates

Akihiro Tomura,[1] Makoto Nomura,[1] Chiaki Ohae,[1,2] and Masayuki Katsuragawa[1,2,*]

[1] *Graduate School of Informatics and Engineering, University of Electro-Communications, 1-5-1 Chofugaoka, Chofu, Tokyo 182-8585, Japan.*
[2] *Institution for Advanced Science, University of Electro-Communications, 1-5-1 Chofugaoka, Chofu, Tokyo 182-8585, Japan.*
* katsuragawa@uec.ac.jp

We report an optical method of generating arbitrary polarization states by manipulating the thicknesses of a pair of uniaxial birefringent plates, the optical axes of which are set at a crossing angle of π/4. The method has the remarkable feature of being able to generate a distribution of arbitrary polarization states in a group of highly discrete spectra without spatially separating the individual spectral components. The target polarization-state distribution is obtained as an optimal solution through an exploration. Within a realistic exploration range, a sufficient number of near-optimal solutions are found. This property is also reproduced well by a concise model based on a distribution of exploration points on a Poincaré sphere, showing that the number of near-optimal solutions behaves according to a power law with respect to the number of spectral components of concern. As a typical example of an application, by applying this method to a set of phase-locked highly discrete spectra, we numerically demonstrate the continuous generation of a vector-like optical electric field waveform, the helicity of which is alternated within a single optical cycle in the time domain.

## I. INTRODUCTION

The use of optical technologies to manipulate physical quantities to define an optical wave has always opened up new possibilities in terms of both engineering applications such as information technology and basic sciences related to the control or measurement of material properties. The establishment of a technology to control the carrier envelope phase of an optical wave (optical frequency comb) [1] has brought about revolutionary developments in optical frequency standards followed by high-resolution laser spectroscopy [2], as well as in its counterpart, attosecond science [3]; this is likely the most symbolic example in a recent development of optical science. Other developments include programmable waveform generation [4–7] by employing a spatial light modulator, time-dependent polarization control [8–11], pulse shaping with metasurfaces [12], controlling terahertz waveforms [13–15], and the use of these technologies to control molecular ionization [9, 10], photocurrent in solids [16–18], magnetization vectors [19, 20], and tunneling currents [21, 22].

Here, we describe an optical technology for generating arbitrary polarization states by manipulating the thicknesses of a pair of uniaxial birefringent plates, the optical axes of which are arranged at a crossing angle of π/4. The essential difference from the widely used method employing a pair of λ/2 and λ/4 waveplates is that this new method can generate arbitrary polarization-state distributions in a group of highly discrete spectra without spatially separating them into their individual spectral components. Generating arbitrary optical electric field waveforms "continuously" in the time domain, like a synthesizer, by arbitrarily controlling each amplitude and phase of a group of highly discrete spectra [23, 24], can be the representative technologies of optical wave control. The polarization-manipulation method proposed here can be used to further provide arbitrary polarization distributions. By incorporating this method with the results of our previous study [25], we numerically demonstrate the "continuous" generation of a vector-like electric field waveform, the helicity of which is alternated within a single optical cycle in the time domain.

## II. PRINCIPLES

### A. Definitions

A plane (electric field) wave propagating along the $z$ direction is expressed as

$$\boldsymbol{E}(t,z) = e^{i\omega t}\begin{pmatrix} A_x e^{i(-kz+\phi_x)} \\ A_y e^{i(-kz+\phi_y)} \end{pmatrix}, \quad (1)$$

where $\omega$ is the angular frequency, $k$ is a wave vector, and $A_{x,y}$ and $\phi_{x,y}$ are the amplitude and phase, respectively, along the x- and y-axes, respectively. The polarization state of this electric field is described by a normalized Jones vector as

$$\boldsymbol{J} = \begin{pmatrix} \cos\chi \\ \sin\chi\, e^{i\delta} \end{pmatrix} \quad (2)$$

$$\chi = \tan^{-1}\frac{A_y}{A_x},\ \delta = \phi_y - \phi_x,$$

where $\chi$ and $\delta$ are the azimuth angle and the relative phase retardance, respectively. This polarization state is expressed by a Stokes vector on the Poincaré sphere as

$$\boldsymbol{S} = \begin{pmatrix} S_1 \\ S_2 \\ S_3 \end{pmatrix} = \begin{pmatrix} \langle A_x^2 - A_y^2 \rangle \\ 2\langle A_x A_y \cos\delta \rangle \\ 2\langle A_x A_y \sin\delta \rangle \end{pmatrix}, \quad (3)$$

where the angle brackets denote the average over time.

### B. Standard method: manipulation of polarization states by using a pair of λ/4 and λ/2 waveplates

As is well known, arbitrary polarization states can be generated by employing a pair of λ/4 and λ/2 waveplates [QWP (quarter-wave plate) and HWP (half-wave plate), respectively] (Fig. 1(a)). Let us assume that the incident light is linearly polarized along the $y$-axis, i.e., $\boldsymbol{J}_{in} = (0,1)^T$. The in-plane rotation angles of QWP and HWP are described as $\psi_1$ and $\psi_2$. The polarization state of the light after passing through these two waveplates is given as

$$\boldsymbol{J}_{out} = \boldsymbol{R}(-\psi_2)\boldsymbol{J}_{wp}(\pi)\boldsymbol{R}(\psi_2 - \psi_1)\boldsymbol{J}_{wp}\left(\frac{\pi}{2}\right)\boldsymbol{R}(\psi_1)\boldsymbol{J}_{in},$$

[26], where $\boldsymbol{R}$ is a rotation matrix and $\boldsymbol{J}_{wp}$ is a Jones matrix of the waveplate:

$$\boldsymbol{J}_{\mathrm{wp}}(\Gamma) = \begin{pmatrix} e^{-i\frac{\Gamma}{2}} & 0 \\ 0 & e^{i\frac{\Gamma}{2}} \end{pmatrix}.$$

$\Gamma$ is the relative phase retardance, defined as $\Gamma = (n_e - n_o)\omega d/c$, where $d$ is the thickness of the plate, $n_o$ and $n_e$ are the ordinary and extraordinary refractive indices, respectively, and $c$ is the speed of light in a vacuum. The Stokes vector notation of $\boldsymbol{J}_{out}$ is given as

$$\boldsymbol{S} = \begin{pmatrix} -\cos 2\psi_1 \cos(4\psi_2 - 2\psi_1) \\ -\cos 2\psi_1 \sin(4\psi_2 - 2\psi_1) \\ -\sin 2\psi_1 \end{pmatrix}. \quad (4)$$

As $\psi_1$ and $\psi_2$ are independent of each other, $2\psi_1$ and $4\psi_2 - 2\psi_1$ vary independently in the range of 0 to $2\pi$. Thereby, Eq. (4)

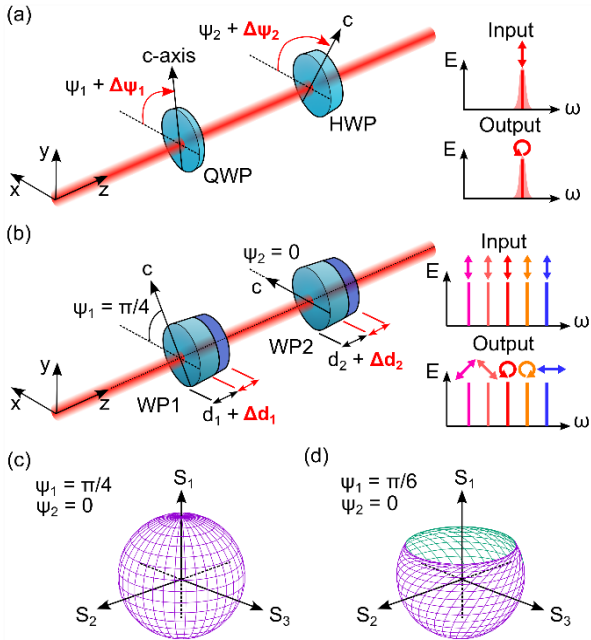

FIG. 1. Two different schemes of generating arbitrary polarization states. (a) Conventional method with a pair of an HWP and a QWP with variable angles of $\psi_1$ and $\psi_2$, respectively. (b) The proposed method with two birefringent plates (WP1 and WP2) of variable thicknesses, $d_1$ and $d_2$, respectively, with a crossing angle, $\psi_1 - \psi_2$, fixed at $\pi/4$. (c) The Poincaré sphere is entirely covered by the Stokes vector when $\psi_1 = \pm\pi/4$ and $\psi_2 = \psi_1 \pm \pi/4$ in scheme (b). (d) The Poincaré sphere is imperfectly covered when $\psi_1 = \pi/6$ and $\psi_2 = 0$.

can be the expression of a Poincaré sphere itself in the polar coordinate system. In other words, by rotating the two waveplates independently, it is possible to generate an arbitrary polarization state.

### C. Proposed method: manipulation of polarization states by using a pair of uniaxial birefringent plates of variable thicknesses

There may be other ways to generate an arbitrary polarization state. A pair of uniaxial birefringent plates (WP1 and WP2) are arranged so that their optical axes make an angle of $\pi/4$ with each other (here, $\psi_1=\pi/4$, $\psi_2=0$). Instead of the in-plane angle, $\psi_{1,2}$, being manipulated, the thickness of each birefringent plate is manipulated independently (Fig. 1(b)). It is possible to generate an arbitrary polarization state in this manner, as shown below.
The incident electric field is assumed to be y-linearly polarized. Then, the polarization state of the output light manipulated in the above manner is given as

$$J_{\text{out}} = R(0) J_{\text{wp}}(\Gamma_2) R(0) R\left(-\frac{\pi}{4}\right) J_{\text{wp}}(\Gamma_1) R\left(\frac{\pi}{4}\right) J_{\text{in}}$$

$$= \begin{pmatrix} \sin\frac{\Gamma_1}{2} \\ \cos\frac{\Gamma_1}{2} e^{i\left(\Gamma_2 + \frac{\pi}{2}\right)} \end{pmatrix}$$

where $\Gamma_{1,2}$ is the relative phase retardance at each of WP1 and WP2, respectively. Then, the Stokes vector representing this polarization state is given as

$$S = \begin{pmatrix} -\cos\Gamma_1 \\ -\sin\Gamma_1 \sin\Gamma_2 \\ \sin\Gamma_1 \cos\Gamma_2 \end{pmatrix}$$

This is also an expression of the Poincaré sphere in the polar coordinates, with $\Gamma_1$ and $\Gamma_2$ as variables. In other words, when the relative phase retardances, $\Gamma_1$ and $\Gamma_2$, are independently manipulated in the range of $[0, \pi]$ and $[0, 2\pi)$, respectively, the polarization of the output light covers the entire Poincaré sphere, generating arbitrary polarization states (Fig. 1(c)). Hereafter, we describe this method in detail.

### D. Features of the proposed method

A remarkable feature of this method (Fig. 1(b)) is its ability to generate arbitrary polarization-state distributions in a group of highly discrete spectra (Fig. 1(b), right panel). Although the standard method (combining the in-plane rotational operations of QWP and HWP) can be used to create one specific polarization state in a (generally continuous) spectrum (Fig. 1(a), right panel), it is generally impossible to create arbitrary distributions of polarization states in the spectrum. The key mechanism of the method (Fig. 1(b)) is that the polarization has a periodicity of $2\pi$ with respect to the thickness of the uniaxial birefringent plate, and this periodicity differs significantly among a group of highly discrete spectra. Despite the manipulation of only a single variable (plate thickness, $d_{1,2}$), a variety of polarization-state distributions are tested (typically over hundreds of periods within a few tens of millimeters of thickness) in a group of highly discrete spectra. The essential difference between the standard method (Fig. 1(a)) and the proposed method (Fig. 1(b)) lies in the mechanism, namely that the former, in principle, limits the range of manipulation to a single rotation whereas the latter does not. In contrast, this method does not work well when an exact solution is pursued. Another key of this method is that, in reality, near-optimal solutions are useful for a variety of applications, and many such solutions can be found in a realistic exploration range.

### E. Requirements of the proposed method

*Requirement 1: Installation angle of the pair of uniaxial birefringent plates*
Before proceeding, we will add a few more words about the requirements for generating arbitrary polarization states by using the proposed method. In order for the output polarization states to cover the entire Poincaré sphere, specific conditions are imposed on the polarization of the incident light and the angles of the optical axes of WP1 and WP2 ($\psi_1$ and $\psi_2$, respectively). After manipulation by using this method, the Jones vector, $J'_{\text{out}}$, in the coordinates of the second uniaxial birefringent plate, WP2, is notated as

$$J'_{\text{out}} = J_{\text{wp}}(\Gamma_2) R(\psi_2 - \psi_1) J_{\text{wp}}(\Gamma_1) R(\psi_1) J_{\text{in}}$$
$$= \begin{pmatrix} e^{-i\Gamma_2} \left[ e^{+i\frac{\Gamma_1}{2}} \cos\psi_1 \sin\psi_3 + e^{-i\frac{\Gamma_1}{2}} \cos\psi_3 \sin\psi_1 \right] \\ e^{+i\frac{\Gamma_1}{2}} \cos\psi_1 \cos\psi_3 + e^{-i\frac{\Gamma_1}{2}} \sin\psi_3 \sin\psi_3 \end{pmatrix}$$

where $\psi_2 - \psi_1 = \psi_3$. The first parameter, $S_1$, of the Stokes vector representing the output polarization state, $J'_{\text{out}}$, is
$$S_1 = \sin 2\psi_1 \sin 2\psi_3 \cos\Gamma_1 - \cos 2\psi_1 \cos 2\psi_3$$
To cover all the polarization states, $S_1$ must at least vary in its full range, *i.e.*, $-1 \leq S_1 \leq 1$. This can be satisfied only when the coefficient of the first term of $S_1$, $\sin 2\psi_1 \sin 2\psi_3$, amounts to $\pm 1$. Therefore, $\psi_1$ and $\psi_2$ must be
$$\psi_1 = \pm\frac{\pi}{4}, \psi_2 = \psi_1 \pm \frac{\pi}{4}.$$
When $\psi_1$ and $\psi_2$ are set under these conditions, the entire Poincaré sphere is covered (Fig. 1(c)). Otherwise, for example, if $\psi_1 = \pi/6$ and $\psi_2 = 0$, the output Stokes vector forms an incomplete Poincaré sphere (Fig. 1(d)).

*Requirement 2: Polarization state of incident light*
As described in the previous subsection, the proposed method imposes the requirement that the incident light is linearly polarized for all the spectral components, at a crossing angle of $\pi/4$ with respect to the optical axis of the first uniaxial birefringent plate, WP1. This restriction on the incident light does not spoil the generality of the method. This is because controlling all the physical quantities that characterize an optical wave (amplitude, polarization, and phase) provides the most arbitrary control of the optical field, and then the manipulation of each quantity, in general, is performed in the following order: amplitude (the pair of HWP and polarizer), polarization (anisotropic transparent medium) as discussed in [25], and phase (isotropic transparent medium). In the proposed method, the assumption that the optical wave is linearly polarized before the polarization manipulation—in other words, linearly polarized by a polarizer after manipulation of the amplitude—does not limit the arbitrariness of the optical wave manipulation. More generally, even if the incident optical wave has an arbitrary polarization-state distribution, it can be transformed to any polarization-state distribution by the installation of one more uniaxial birefringent plate before the pair of birefringent plates, WP1 and WP2 (see Section A in [27]).

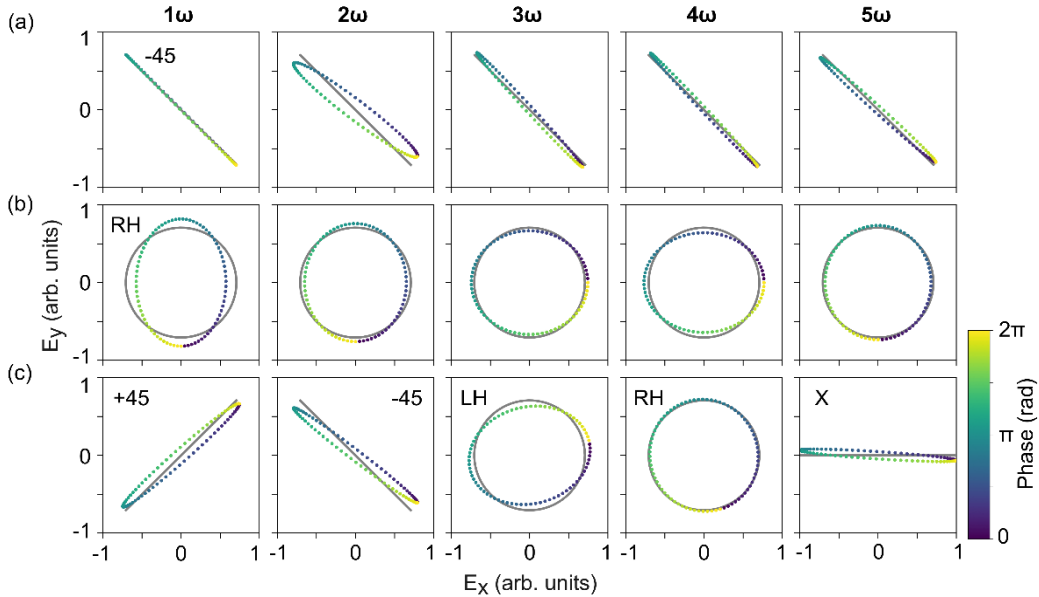

FIG. 2. Traces of Jones vectors obtained as optimal solutions for three targets. (a) −45 degrees linear polarization; (b) RH (right-handed) circular polarization; (c) Mixed target: +45 degrees linear, −45 degrees linear, LH (left-handed) circular, RH (right-handed) circular, and X linear for 1 to 5 w. For comparison, the corresponding target is also plotted in each panel (solid gray line).

### III. RESULTS AND DISCUSSION: NUMERICAL EXPERIMENTS

#### A. Generation of arbitrary polarization-state distributions in a group of five highly discrete spectra

Here, we show the results of numerical experiments in which the method of arbitrarily manipulating polarization states, as described in Section II. C (Principles), was applied in a realistic situation. We employed a group of five highly discrete spectra, each of which had an integer multiple frequency of 125 THz, extending from the near infrared to the visible wavelength region ($\omega$: 125 THz, 2400 nm; $2\omega$: 250 THz, 1200 nm; $3\omega$: 375 THz, 800 nm; $4\omega$: 500 THz, 600 nm; $5\omega$: 625 THz, 480 nm). Here, we assumed crystal quartz as the material of the uniaxial birefringent plates [28] (see Section B in [27]). We placed a pair of crystal quartz plates (WP1 and WP2) coaxially on the optical axis (as illustrated in Fig. 1(b)), and we varied each of the plate thicknesses, $d_1$ and $d_2$, up to 50 mm with a step size of 0.1 μm. The polarization state of the incident light was set to $y$-linear polarization for all the spectral components (see Section II. E *Requirement 2*). We then set a variety of polarization-state distributions as targets and numerically explored the optimal solutions. As an error function to evaluate deviation from the targets, we employed the mean-squared (MS) (Euclidean) distances (see Section C in [27]) for the five spectral polarization states. This was a numerical demonstration of the arbitrary manipulation of polarization states to be used in line with the arbitrary optical-waveform generation studied in [25].

We plotted the optimal solutions obtained within the exploration range for three different targets (Fig. 2). We visualized them by the one-period behaviors of the Jones vectors. In Figs. 2(a) and 2(b), we set a single polarization state as a target for all five components, namely (a), –45 degrees linear polarization and (b), right-handed circular polarization. To more clearly demonstrate the capacity of this method, in Fig. 2(c) we set a more random target, namely +45 degrees linear, –45 degrees linear, left-handed circular, right-handed circular, and $x$ linear for each of $1\omega$ to $5\omega$. Below, for simplicity, ± 45 degrees of linear polarization, left-handed/right-handed circular polarization, and $x/y$ linear polarization are denoted as ± 45, LH/RH, and X/Y, respectively. As a result of our explorations, we achieved good approximate solutions for each of the three different targets in Fig. 2(a), 2(b), and 2(c).

#### B. Distribution of near-optimal solutions

We plotted the observed deviations from the targets around the optimal solution obtained, where $\Delta d_1$ and $\Delta d_2$ indicate the thickness changes of WP1 and WP2, respectively, with respect to the optimal position (Fig. 3(a) to 3(c)). The range of 0.8 mm is plotted for each of $\Delta d_1$ and $\Delta d_2$. This range corresponds to about 0.03% of the entire exploration area. See Section D in [27] for the error behaviors over a wider area. The horizontal and vertical axes, $\Delta d_1$ and $\Delta d_2$, correspond to the circumferential motions on the Poincaré sphere (Fig. 3(d)); the former is on the cross-section cut out by the plane containing the axis $S_1$ (the latitude), and the latter is on the cross-section cut out by the plane parallel to $S_2$–$S_3$ (the longitude). Each Stokes vector of the five spectral components moves at a markedly different speed on the Poincaré sphere, forming the interference lattice patterns of the sum of the MS errors from the respective target polarization states (Fig. 3(a) to 3(c)). From a perspective viewpoint, the interference lattice patterns gradually collapse with about a 0.2-mm thickness change, followed by the revival of similar lattice patterns. The key property of this method is concisely determined by these characteristic behaviors, namely, we cannot know exactly where an optimal solution will appear, but we can predict how frequently allowable near-optimal solutions may appear if we explore over a certain range. Additionally, deviations from the targets tend to be small along a certain path where the ratio of the two thicknesses $\Delta d_1$ and $\Delta d_2$ is approximately constant (the red slopes in Fig. 3(a) to 3(c)). It is possible efficiently to explore the near-optimal solutions by tracing these paths.

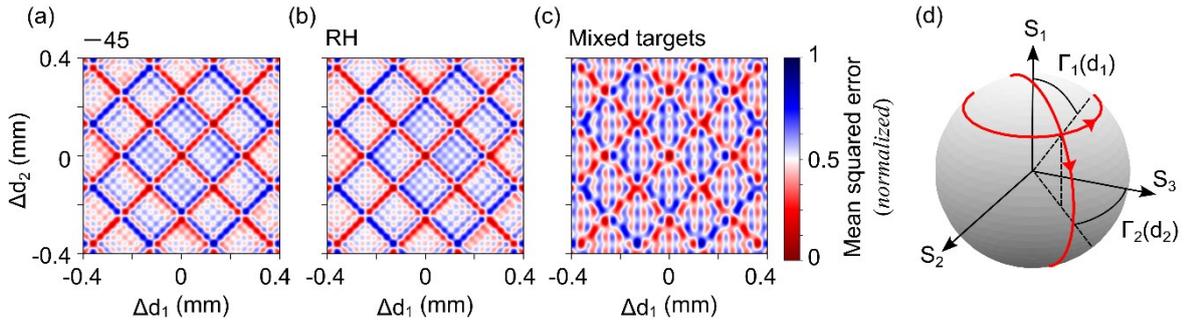

FIG. 3. Two-dimensional maps of deviation from the targets plotted as functions of the birefringent plate thicknesses, $d_1$ and $d_2$. (a) –45 degrees linear; (b) RH circular; (c), Mixed target. The center of each map corresponds to the optimal solution obtained. (d) Poincaré sphere representation of the Stokes vector; the red curves illustrate traces corresponding to scanning of the plate thickness $d_1$ or $d_2$.

| Targets | –45 degrees linear | RH circular | Mixed target |
|---|---|---|---|
| SD (deg.) / (%) | 10.61 / 5.89 | 11.31 / 6.28 | 11.58 / 6.43 |
| Projection | 0.983 | 0.980 | 0.980 |
| Number of solutions | 11675 | 5575 | 10757 |
| Probability ($10^{-10}$) | 467 | 223 | 430 |

TABLE I. Evaluation of near-optimal solutions. Deviations from the targets are evaluated and represented as spherical distance (SD) (first row), and also as projections to the target vector (second row). The third row indicates the number of near-optimal solutions where the normalized MS error (Euclidian distance) reaches less than 0.02. The bottom row shows the probabilities of finding a near-optimal solution.

Table I summarizes the evaluations of the near-optimal solutions found in our numerical explorations. It includes the number of near-optimal solutions within an allowable error (MS error < 0.02) and the inferred probabilities of finding a near-optimal solution. Deviations from the targets are evaluated either by the spherical distance (SD) or by the projection of an optimal solution onto the target (see Section C in [27]). The table shows that the optimal solutions achieved approximately 98% similarity by the projection to the targets and about 6% deviation from the targets by the spherical distance. As already mentioned, this method does not function if an exact solution is pursued (see Section E in [27]). By introducing an acceptable error, depending on the aim, one can indeed find a sufficient number of near-optimal solutions within a realistic exploration range (a few tens of millimeters), as exemplified in Table I.

In this experimental section, we demonstrated a numerical experiment assuming actual frequencies and materials. We showed that, on the basis of the method presented in Section II. C to E, it is possible to generate arbitrary polarization-state distributions in a group of five highly discrete spectra without spatially separating them. Synthesis of optical waves is generally an extremely high hurdle once each of the spectral components is spatially separated, because both the optical paths and the spatial modes must be exactly matched, with a precision of the optical phase, among all the spectral components over a long period, as can be seen in recent studies of the coherent addition of laser beams [29]. The arbitrary polarization-manipulation technology presented here, which does not spatially separate a group of spectral components, has a great advantage in practical applications.

## IV. DISCUSSION: DETAILED PROPERTIES AND DISTRIBUTION OF NEAR-OPTIMAL SOLUTIONS

In this section, we discuss the properties of the near-optimal solutions in more detail, namely how they are distributed on the Poincaré sphere and how they behave differently depending on the given conditions.

Deviations from a target form a certain probability distribution, which can be reproduced and interpreted well on the basis of the concise model proposed in this section. Let us consider the case where the number of spectral components, N, equals 1. Polarization states having an equal deviation, $s \pm \Delta s$, from a given target form a circular band on the Poincaré sphere. If we take a sufficient number of exploration points on the Poincaré sphere into consideration, then the ratio of the number of points within this band to the total number of exploration points gives the probability of finding the polarization state having a deviation within $s \pm \Delta s$.

The distribution of the exploration points depends on how the thicknesses of the birefringent plates are manipulated. Here, we varied the thicknesses of the plates WP1 and WP2, i.e., $d_1$ and $d_2$, at a constant increment. In this case, the exploration points are distributed on the Poincaré sphere such that the angle variables $\Gamma_1$ and $\Gamma_2$ change with equal spacing. Therefore, in this operational method, a greater number of exploration points are distributed around the north and south poles ($\Gamma_1 = m \pi$, where m is a natural number) and, conversely, a smaller number are distributed around the equator ($\Gamma_1 = \pi /2 + m \pi$). Consequently, the dependence on the location of a target on the Poincaré sphere is incorporated into the probability distribution of the deviation from the target.

In the case of multiple and discrete spectral components (N ≥ 2), deviation from a given target (the polarization-state distribution of multiple spectral components) has to consider all the contributions of the deviations of the spectral polarization states from the targets. If we assume that the deviation distribution does not depend on the frequency of the spectral component under the manipulation, and that each of the polarization states behaves independently on the Poincaré sphere, then the distribution of the total deviation of multiple spectral components can be given by convolving each of the deviation distributions sequentially by N − 1 times. The total deviation distribution calculated by using this model reproduces well the behavior obtained in the numerical experiments, including the tail regions of the distributions with small probabilities, which provide near-optimal solutions.

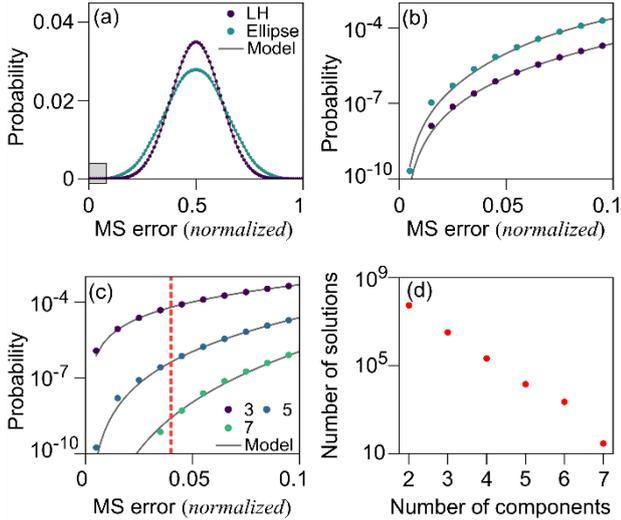

FIG. 4. Detailed properties of near-optimal solutions. (a) Entire behaviors of the probability distribution, including dependence on the targets: LH and ellipse. N = 5; exploration range, 100 x 100 mm$^2$. (b) Extended view of the left tail region of (a) (shadowed region). (c) Dependence of the probability of near-optimal solutions on the number of spectral components, N. Target, LH; exploration range, 100 x 100 mm$^2$. (d) Number of near-optimal solutions as a function of N.

In Fig. 4(a), two typical examples of the deviation distributions observed in the numerical experiment for the group of five spectral components examined in Section III are exhibited (colored dots); one corresponds to the target, LH (located at the equator, $\Gamma_1 = \pi/2$), and the other to an elliptical polarization (Ellipse; at mid-latitude, $\Gamma_1 = \pi/4$). The horizontal and vertical axes correspond to the deviation (MS error) normalized by its maximum value and the probability, respectively. Modeled deviation distributions are shown by the gray solid lines. The deviation distributions predicted by the above concise model are in good agreement with those observed in the numerical explorations, including the target dependence.

Figure 4(b) is an extended view of the left tail region of the distribution profiles in Fig. 4(a), providing near-optimal solutions. The target dependence appears strongly in this region of small deviation. As mentioned above, a greater number of exploration points are distributed around the poles. Therefore, the probability of finding a near-optimal solution is expected to increase when the target is located near the poles. It turns out that the model indeed predicts the deviation distributions, including such a small probability region, providing near-optimal solutions. Figure 4(b) also shows that the probability increases or decreases nearly exponentially with respect to the deviation, thus, the number of near-optimal solutions increases dramatically when a slightly larger error tolerance is provided, and vice versa. This property can be attributed to the mechanism that each of the polarization states in a group of multispectral components behaves almost independently; thereby, the behavior of the total deviation from the target is given as the convolution of each of the deviation distribution.

A realistically controllable number of discrete spectral components can also be estimated from this property of the total deviation. The probabilities of the near-optimal solutions decrease with a power law with respect to a given number, N, of spectral components to be controlled (Fig. 4(c) and 4(d)). In reality, the number of spectral components to be simultaneously controlled is determined on the basis of this property; this includes the controllable thicknesses of the birefringent plates and the amount of refractive index dispersion among the spectral components.

In this Discussion section, we have described how the distribution of the deviation of polarization states from a given target is reproduced well by a concise model, and how, on the basis of this model, we can infer enough of the information required to apply this arbitrary polarization-manipulation method in reality. This information includes the distribution properties of the near-optimal solutions; the exploration ranges required to obtain a near-optimal solution with the requested accuracy; and the number of simultaneously controllable spectral components.

## V. APPLICATION: CONTINUOUS GENERATION OF VECTORIAL ELECTRIC FIELD WAVEFORMS

Lastly, we present a numerical demonstration in which the proposed method of arbitrary polarization manipulation is used to control an optical electric field waveform in the time domain. The optical technology of continuously generating arbitrary electric field waveforms, just like a synthesizer in electronics, can be one of the representative technologies in optical wave control. As theoretically and experimentally discussed in [24, 25, 30], manipulation of the amplitude and phase of each of five phase-locked spectra having an exact integer frequency ratio can continuously produce arbitrary electric field waveforms in the time domain. The method proposed here appends an extra degree of freedom, namely the arbitrary manipulation of the polarization-state distribution in a group of such highly discrete spectra.

We show a typical example in Fig. 5(a), 5(b), and 5(c). After we manipulate the polarization states of 1, 3, and 5 ω to x-linear polarization and those of 2 and 4 ω to y-linear polarization [15] by applying the proposed method, the spectral phases of 1 to 5 ω are set to π/2, 0, π/2, 0, and π/2, respectively (Fig. 5(a)). The dotted line in Fig. 5(b) shows the electric field waveform retrieved by the achieved polarization-state distribution, in which the helicity is alternated in a single optical cycle. Figure 5(c) compares the waveform achieved in Fig. 5(b) with the target. It can be seen that the achieved polarization-state distribution provides a satisfactory solution in reality.

## VI. CONCLUSIONS

Here, we have described an optical technology for generating arbitrary polarization states by manipulating the thicknesses of each of a pair of uniaxial birefringent plates, the optical axes of which are arranged in relation to each other at a crossing angle of π/4. The essential difference from the widely used method of manipulation of a pair of λ/2 and λ/4 waveplates is that this new method can generate an arbitrary polarization-state distribution in a group of highly discrete spectra without spatially separating them into their individual spectral components. Through an exploration, we have shown that the target polarization-state distribution can be obtained as one of the near-optimal solutions and that a sufficient number of near-optimal solutions is found within a realistic exploration range. We have also shown that the properties of such near-optimal solutions, including the exploration range required to find them and the number of controllable spectral components, are described well by a concise model on Poincaré spheres. As a typical example of application, we have numerically demonstrated the continuous generation of a vectorial optical electric field waveform in the time domain, the helicity of which alternates in a single optical cycle, by applying the proposed method to a group of five highly discrete phase-locked spectra and using the arbitrary manipulation of amplitudes and phases investigated in [25]. This method of arbitrarily manipulating polarization states can be regarded as an optical technology that provides a new degree of freedom in material control or information processing involving light.

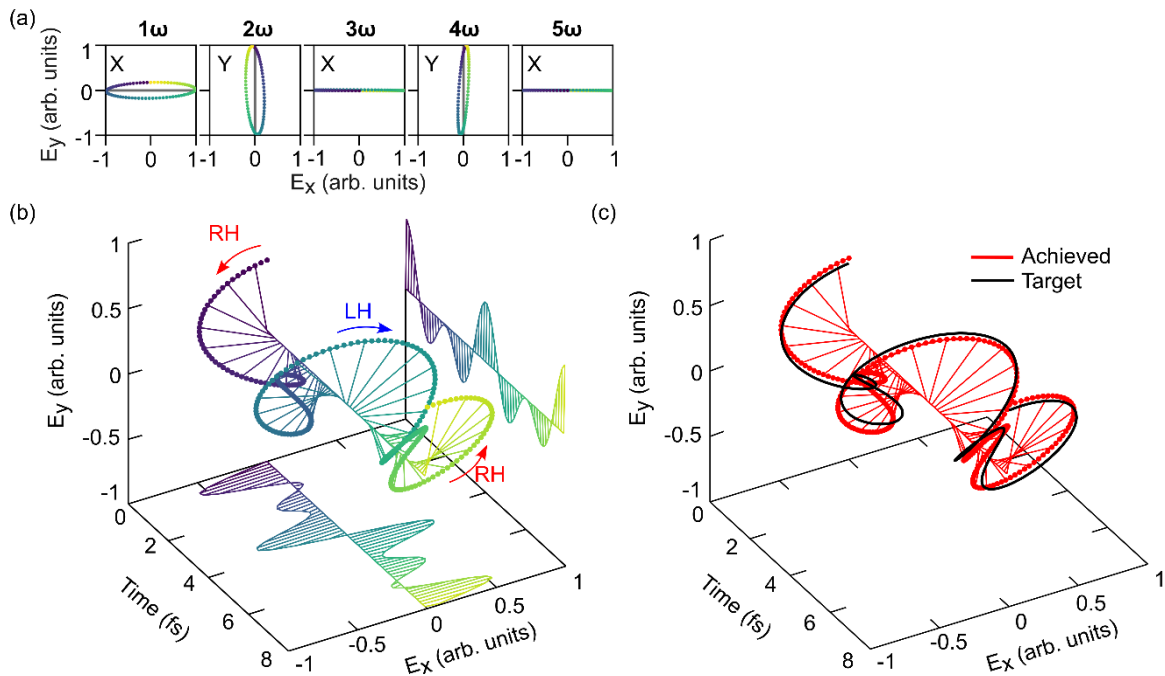

FIG. 5. Numerical demonstration of full vectorial control of an electric field waveform under a continuous generation regime. (a) The optimal solutions found in the exploration of the target, namely X, Y, X, Y, and X for 1 to 5 ω, respectively. The spectral phase is assumed to be controlled by the method discussed in [25] and has the values π/2, 0, π/2, 0, and π/2 for 1 to 5 ω, respectively. (b) An ultrafast waveform retrieved by using the polarization state achieved in (a). (c) Comparison between the optimal waveform achieved in (b) and the target.

ACKNOWLEDGMENTS

This work was supported by Grants-in-Aid for Scientific Research (S) No. 20H05642, (A) No. 24244065, and (B) No. 20H01837, JST PRESTO JPMJPR2105.

*Email address: katsuragawa@uec.ac.jp